\documentclass[aps,prl,twocolumn,showpacs,groupedaddress,amsmath,amssymb]{revtex4}

\usepackage[dvips]{graphicx}
\usepackage{bm}
\begin{document}


\title{Self Running Droplet: Emergence of Regular Motion from Nonequilibrium Noise}


\author{Yutaka Sumino}
\affiliation{Depertment of Physics, Graduate School of Science, Kyoto University \& CREST, Kyoto, 606-8502, Japan}

\author{Nobuyuki Magome}
\altaffiliation{Present address: Department of Food and Nutrition, Nagoya Bunri College, Nagoya, 451-0077, Japan.}
\affiliation{Depertment of Physics, Graduate School of Science, Kyoto University \& CREST, Kyoto, 606-8502, Japan}

\author{Tsutomu Hamada}
\affiliation{Depertment of Physics, Graduate School of Science, Kyoto University \& CREST, Kyoto, 606-8502, Japan}

\author{Kenichi Yoshikawa}
\email[To whom correspondance should be addressed. Tel:+81-75-753-3812. Fax:+81-75-753-3779. Email:]{yoshikaw@scphys.kyoto-u.ac.jp}
\affiliation{Depertment of Physics, Graduate School of Science, Kyoto University \& CREST, Kyoto, 606-8502, Japan}


\date{\today}

\begin{abstract}
Spontaneous motion of an oil droplet driven by chemical nonequilibricity is reported. It is shown that the droplet undergoes regular rhythmic motion under appropriately designed boundary conditions, whereas it exhibits random motion in an isotropic environment. This study is a novel manifestation on the direct energy transformation of chemical energy into regular spatial-motion under isothermal conditions. A simple mathematical equation including noise reproduces the essential feature of the transition from irregularity into periodic regular motion. Our results will inspire the theoretical study on the mechanism of molecular motors in living matter, working under significant influence of thermal fluctuation.
\end{abstract}


\maketitle

The second law of thermodynamics prohibits specific spontaneous motion in macroscopic systems under equilibrium; nature obeys the law of equipartition. In contrast, living organisms exhibit a specific spatio-temporal structure in a self-organized manner under thermodynamically open conditions, where chemical processes play the major role. Over the past several decades, reaction-diffusion systems have attracted considerable attention as a simple model for the formation of spatio-temporal structure in living organisms. It has been shown both experimentally and theoretically that various gdissipative structuresh can be generated, such as travelling waves and a Turing pattern~\cite{selforgnprig,mathbio,kuramoto97}. However, in a reaction-diffusion system there is essentially no mass-flow, or geometric motion, in contrast to actual observations in living things. In the present study, we found regular spatial movement could be induced in a greactive droplethby adopting an appropriate boundary condition under chemically nonequilibrium conditions. We will discuss a novel scenario for the generation of macroscopic directed and/or periodic motion under a large fluctuation.

Since the 19th century, it has been known that an oil/water system exhibits spontaneous mechanical agitation, i.e. the Marangoni effect~\cite{dupeyrat78,takahasi02,kovalchulk04,magome96,yoshikawa_omochi86,shioi03}. Two different kinds of self-agitation, thermal and chemical Marangoni convection, have been reported. Chemical Marangoni convection is described as the time-dependent fluctuation of interfacial tension under isothermal conditions, where the driving force is the transfer of solute through the interface. One type of chemical Marangoni convection is the spontaneous mechanical motion of a reactive liquid droplet on a solid substrate. The motion of a reactive droplet has been predicted theoretically~\cite{greenspan78,degenne85,brochard89,degennecap04,degenne98}and verified experimentally~\cite{bain89,chaudhury92,bain94,santos95,slee00,bico00,slee02}. However, in such a system the trajectory of motion cannot overlap itself because the substrate surface is changed in an irreversible manner, which would preclude periodic or circular motion. In the present study we found that irregular agitation induced by a chemical Marangoni effect~\cite{dupeyrat78, magome96,shioi03} can be transduced into characteristic repetitive, rhythmic motion, simply by choosing appropriate boundary conditions.

The aqueous phase contained 1 mM stearyl trimethyl ammonium chloride, and the organic phase was 5mM iodine solution of nitrobenzene saturated with potassium iodide are used. Stearyl trimethyl ammonium chloride was prepared through recrystallization from acetone. For glass substrates, we used micro slide glass (Matsunami, Osaka; S9111) without any pre-treatment. As for a measurement $20-60 \mu$l of the oil is placed in an aqueous phase on a glass substrate. Movement of the oil droplet is recorded by high speed video camera (RedLake MASD Inc.) Motion Scope PCI in 60 frames per second. All the measurements are carried out at the room temperature.

\begin{figure}
\includegraphics[width=7cm]{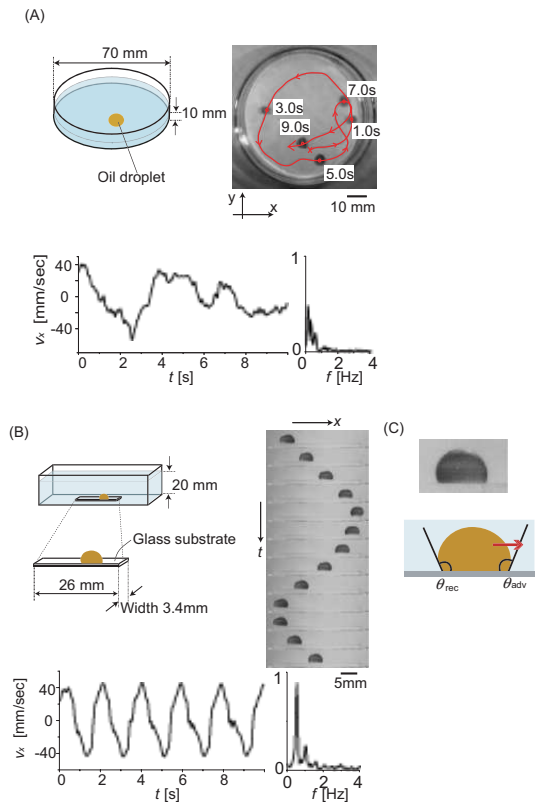}
\caption{According to the configuration of the glass substrate, an oil droplet showed a distinct mode of motion. (A): (upper left) Schematic diagram of the experimental setup. The volume of the oil droplet is $30 \mu$l. (upper right) Random motion appears under isotropic conditions. Also see movies in~\cite{movies}. (lower figure) The trace of the $x$-component of oil-droplet velocity and its Fourier transformation, where there is no distinct peak. The $y$-component velocity shows the similar behaviour. (B): (upper left) Schematic diagram of the experimental setup. The volume of the oil droplet is $30\mu$l. (upper right) Periodic forward-and-backward motion appears on a narrow and straight glass substrate. The time scale in the picture on the right is 1/6 second per frame. The rhythmic motion continues for 10 seconds. Also see movies in~\cite{movies}. (lower figure) Trace of the magnitude of oil-droplet velocity and its Fourier transformation, indicating a   peak at 0.5 [Hz]. (C): Schematic representation of an oil droplet, moving toward the red arrow. $\theta_{\mathrm{adv}}$ was smaller than $\theta_{\mathrm{rec}}$. On average, $\theta_{\mathrm{adv}} \approx 105^{\circ}$, $\theta_{\mathrm{rec}} \approx 116^{\circ}$.}
\label{fig1}
\end{figure}

Figure \ref{fig1} shows the motion of an oil droplet on a glass substrate in an aqueous phase. We put a small amount (30$\mu$l) of oil on the glass substrate through a pipette. The oil droplet spontaneously begins to move immediately after it is transferred to the glass substrate. As shown in Fig.\ref{fig1}A, the oil droplet exhibits random agitation within the circular glass vessel. Note that the trajectory of the oil-droplet motion crosses itself. In contrast to the irregular motion shown in Fig.\ref{fig1}A, a regular rhythmic motion can be generated in the same chemical system simply by changing the spatial geometry, as in Fig.\ref{fig1}B. Corresponding movies are available online ~\cite{movies}. Such periodic motion along the glass substrate continues over several tens of seconds and then stops. Figure \ref{fig1} also shows time traces of the $x$-component of velocity, together with their Fourier transformation, indicating the appearance of periodicity $f \approx 0.5$ Hz for motion on the glass strip (B) in contrast to the noisy behaviour in the circular vessel (A). Careful observation of Fig.\ref{fig1}B reveals that the advancing contact angle of the oil droplet is smaller than its receding contact angle. Figure \ref{fig1}C shows the enlarged image of an oil droplet moving towards the right, where the advancing contact angle is greater than its receding angle. This large difference in contact angles suggests that the driving force can be attributed to the difference in surface tension between the front and rear of the oil droplet.

\begin{figure}
\includegraphics{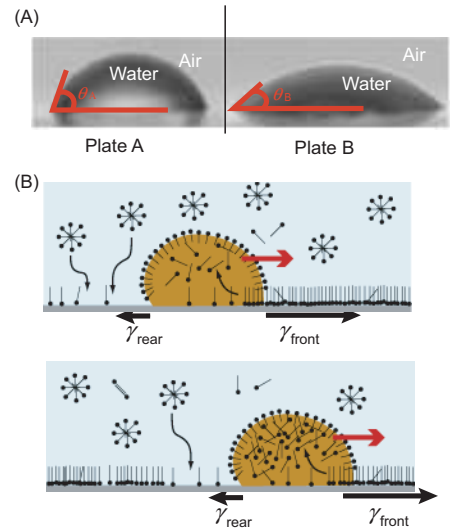}
\caption{(A): A water droplet on a glass substrate in air. We estimate the difference in water-glass surface tension $\Delta\gamma$ from $\Delta\gamma = \gamma_{\mathrm{pure water}} (\cos\theta_{\mathrm{B}}- \cos\theta_{\mathrm{A}})$. Since $\theta_{\mathrm{A}}\approx 68^{\circ}$, $\theta_{\mathrm{B}}\approx 43^{\circ}$, and $\gamma_{\mathrm{pure water}} \approx 72$ [mN/m], the difference in the interfacial tension is $\Delta\gamma \approx 14$ [mN/m]. (B):Schematic diagram of the mechanism of the difference in surface tension. Red arrow shows the direction of oil-droplet motion. STA$^{+}$ is represented as a hydrophobic bar with a polar head. Since STA$^{+}$ tends to assemble with I$_{3}^{-}$ in the droplet, STA$^{+}$ dissolves into the oil droplet from the glass surface. Thus, a difference surface tension is generated as $\gamma_{\mathrm{front}}>\gamma_{\mathrm{rear}}$. The glass surface returns to hydrophobic shortly after an oil droplet passes by, since STA$^{+}$ in the aqueous phase again adheres to the glass surface.}
\label{fig2}
\end{figure}

Next, we conducted an experiment to check for differences in the surface of the glass plate by mimicking the conditions before and after the oil droplet had passed.
We examined the contact angle of a droplet with pure water on the glass plates A and B under air. The glass plate A was pre-treated with the procedure of immersion in an aqueous solution for 5 minutes, rinse with distilled water, and then drying in the air. The other substrate B was pre-treated in the same procedure as in plate A, then dipped into an organic phase for 5 minutes, carefully rinsed with distilled water, and dried in the air.
A large difference in the contact angle was observed [$\theta_{\mathrm{A}} \approx 68^{\circ}$, $\theta_{\mathrm{B}} \approx 43^{\circ}$] (Fig.\ref{fig2}A), indicating that plate B is less hydrophobic than plate A. Thus, it is apparent that a difference in surface tension $\Delta\gamma$ between the front and rear of an oil droplet is generated accompanied by its translational motion. Once $\Delta\gamma$ becomes non-zero due to some spontaneous small motion driven by the Marangoni effect, an oil droplet tends to continue the directed motion by avoiding the backward movement due to the memory effect of the chemical condition \cite{degenne98,slee00}.

The generation of a difference in surface tension between the front and rear can be explained as schematically shown in Fig.\ref{fig2}B. Since the surface of the glass substrate has a negative charge, STA$^{+}$ (Stearyl Trimethyl Ammonium Ion) is arranged on the glass plate with the head group attached to the surface and the hydrophobic tail toward the water phase; as a result, the glass surface that faces the aqueous solution is hydrophobic. When an oil droplet exhibits a directed motion due to the Marangoni effect, STA$^{+}$ attached to the surface tends to dissolve into the organic phase. Consequently, the glass surface becomes less hydrophobic, and the difference in surface tension is generated between the front and rear of the droplet. This imbalance in the surface tension promotes directed motion against viscous damping. The existence of I$_{3}^{-}$ made from I$^{-}$ and I$_{2}$ in an oil droplet accelerates the transfer of STA$^{+}$ from the surface into the organic phase \cite{shioi03} by creating a hydrophobic ion pair with STA$^{+}$. Significantly, the surface of the glass substrate spontaneously returns to the hydrophobic state soon after an oil droplet passes by, since STA$^{+}$ in the aqueous phase tends to be absorbed onto the glass substrate. This mechanism allows the trajectory of an oil droplet to cross itself; thus, rhythmic and repetitive motion is generated in our system.

An oil droplet ceases its motion accompanied with the decrease of the chemical nonequilibricity. This is explained as follows. As I$_{3}^{-}$ concentration decreases in the oil droplet because of pairing with STA$^{+}$, the oil droplet cannot afford enough driving force, then the oil droplet stops. We have also found that with the decrease of the size of droplet, its translational speed tends to be larger and the lifetime of self-motion to be shorter. This may be due to the enhanced surface effect on smaller system.

Let us now discuss the mechanism of the appearance of regular, rhythmic motion in the self-agitating system. The characteristic feature of the motion of an oil droplet can be described as
\begin{equation}
m\frac{d\bm{v}}{dt}=-\eta\bm{v}+\epsilon\frac{\bm{v}}{|\bm{v}|}+\bm{\xi}(t). 
\end{equation}
\begin{figure}
\includegraphics{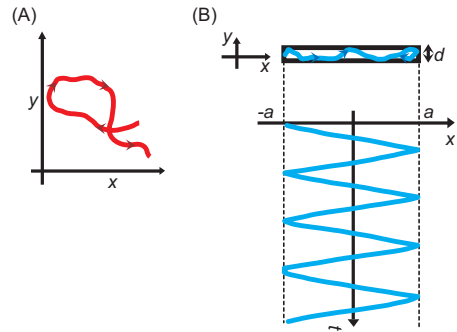}
\caption{Numerical simulation on the spontaneous motion of an oil droplet at different boundary condition~\cite{simulation}. 
(A): Random motion of an oil droplet under boundary-free isotropic environment, which is shown as thetwo-dimensional trace of the motion. (B): Repetitive motion of an oil droplet on a narrow stripe that restricts the motion to $-a < x < a$ and $-d/2 < y <d/2$, where the upper figure shows the trace of the motion in real space and the lower figure shows spatio-temporal representation of the motion. Oil-droplet motion described by Eq.1 varies distinctively depending on the effective freedom of motion that is restricted by a boundary condition. In (B), an oil droplet shows forward and backward motion when $d$ is small enough. 
}
\label{fig3}
\end{figure}
The first term represents the resistance force and the second term represents the driving force caused by the difference in surface tension, where we incorporate the characteristics of self-motion to prohibit backward motion as in Fig.\ref{fig2}B. The third term $\bm{\xi}$ represents the Gaussian random force whose standard deviation is $\sigma$ resulting from the inhomogeneity of the glass surface and the oil/water interface. Figure \ref{fig3} shows schematic examples of simulated oil-droplet motion based on Eq.1~\cite{simulation}. We consider that the driving force is comparable to the Gaussian random force ($\epsilon \sim \sigma$). As for the boundary condition corresponding to the edge of a glass substrate, an oil droplet is assumed to exhibit inelastic rebound at the boundary since there exists dissipation caused by inversing convective flow inside the oil droplet.
In the absence of edge effect on a vessel, random motion is generated(Fig.\ref{fig3}A). In contrast, when the width of the glass substrate $d$ becomes narrower, i.e. when a quasi one-dimensional boundary condition is imposed, periodic regular motion is generated (Fig.\ref{fig3}B)  \cite{one_d}.
This result indicates that the reduction of the effective freedom of motion with an appropriate boundary condition can cause a transition from irregular behaviour to regular behaviour for such a droplet under chemical nonequilibricity.

\begin{figure}
\includegraphics{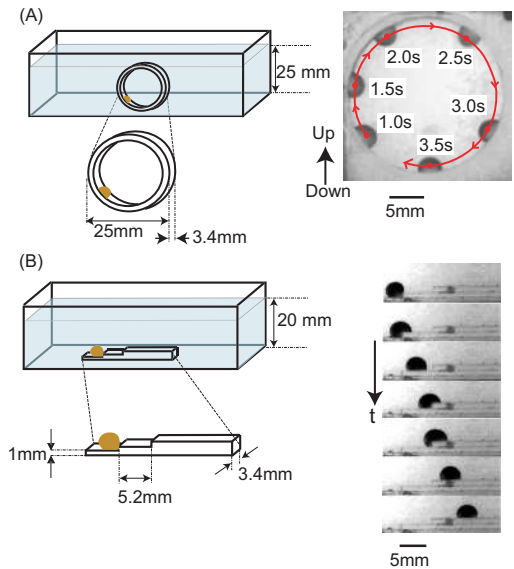}
\caption{(A): Rotational motion within a circle normal to the horizontal plane. In the experiment shown in this figure, the droplet undergoes rotation up to three times and then stops. The volume of the oil droplet is 20$\mu$l. Also see movies in~\cite{movies}. (B): Climbing motion on a stair-like substrate. The oil droplet tends to move up the stair, implying that the stair-like substrate can regulate the direction of motion. The time interval in the pictures on the right is 1/6 second per frame. The volume of the oil droplet is 60$\mu$l. Also see movies in~\cite{movies}.}
\label{fig4}
\end{figure}

An oil droplet shows various types of regular motion depending on the shape of the glass substrate. Figure \ref{fig4}A shows an oil droplet moving around a vertical circle, like a roller coaster \cite{force}. The droplet can move up the circle against gravity, where the density of the oil droplet $\rho_{\mathrm{oil}} \approx 1.2$ and the density of the aqueous phase $\rho_{\mathrm{water}} \approx 1.0$. Figure \ref{fig4}B shows the stepping-up motion of an oil droplet. Based on observations in multiple experiments, we have found that the oil droplet tends to climb up the stairs rather than to fall down. Corresponding movies are available online ~\cite{movies}.

In the present study, we have realized regular motion under nonequilibrium noise by setting suitable boundary conditions, where the broken spatial symmetry in the container of the oil/water system reduces the freedom of motion, inducing regular motion of the droplet. Recent studies on single motor proteins \cite{schliwa03} have indicated that the mode of such mechanical motion can be switched by changing the internal and boundary conditions \cite{endow_higuchi00,okuda_hirokawa00,inoue_yanagida01}. Although the spatial scale of molecular motors in living organisms is much smaller than that of the present experimental system, the generation of regular motion under nonequilibrium in a real experiment may inspire the theoretical consideration of the working mechanism of biological molecular motors.

This work is supported by the Grant-in-Aid for the 21st Century COE ``Center for Diversity and Universality in Physics" from the Ministry of Education, Culture, Sports, Science and Technology (MEXT) of Japan, and the Japan Space Forum.


\begin{thebibliography}{29}
\expandafter\ifx\csname natexlab\endcsname\relax\def\natexlab#1{#1}\fi
\expandafter\ifx\csname bibnamefont\endcsname\relax
  \def\bibnamefont#1{#1}\fi
\expandafter\ifx\csname bibfnamefont\endcsname\relax
  \def\bibfnamefont#1{#1}\fi
\expandafter\ifx\csname citenamefont\endcsname\relax
  \def\citenamefont#1{#1}\fi
\expandafter\ifx\csname url\endcsname\relax
  \def\url#1{\texttt{#1}}\fi
\expandafter\ifx\csname urlprefix\endcsname\relax\def\urlprefix{URL }\fi
\providecommand{\bibinfo}[2]{#2}
\providecommand{\eprint}[2][]{\url{#2}}

\bibitem[{\citenamefont{Nicolis and Prigogine}(1977)}]{selforgnprig}
\bibinfo{author}{\bibfnamefont{G.}~\bibnamefont{Nicolis}} \bibnamefont{and}
  \bibinfo{author}{\bibfnamefont{I.}~\bibnamefont{Prigogine}},
  \emph{\bibinfo{title}{Self Orgainization in Nonequilibrium Systems.}}
  (\bibinfo{publisher}{Wiley}, \bibinfo{year}{1977}).

\bibitem[{\citenamefont{Murray}(1990)}]{mathbio}
\bibinfo{author}{\bibfnamefont{J.~D.} \bibnamefont{Murray}},
  \emph{\bibinfo{title}{Mathematical Biology}}
  (\bibinfo{publisher}{Springer-Verlag}, \bibinfo{year}{1990}).

\bibitem[{\citenamefont{Mori and Kuramoto}(1997)}]{kuramoto97}
\bibinfo{author}{\bibfnamefont{H.}~\bibnamefont{Mori}} \bibnamefont{and}
  \bibinfo{author}{\bibfnamefont{Y.}~\bibnamefont{Kuramoto}},
  \emph{\bibinfo{title}{Dissipative Structures and Chaos}}
  (\bibinfo{publisher}{Springer-Verlag}, \bibinfo{year}{1997}).

\bibitem[{\citenamefont{Dupeyrat and Nakache}(1978)}]{dupeyrat78}
\bibinfo{author}{\bibfnamefont{M.}~\bibnamefont{Dupeyrat}} \bibnamefont{and}
  \bibinfo{author}{\bibfnamefont{E.}~\bibnamefont{Nakache}},
  \bibinfo{journal}{Bioelectrochem. Bioenerg.} \textbf{\bibinfo{volume}{5}},
  \bibinfo{pages}{134} (\bibinfo{year}{1978}).

\bibitem[{\citenamefont{Takahashi et~al.}(2002)\citenamefont{Takahashi, Yui,
  and Sawada}}]{takahasi02}
\bibinfo{author}{\bibfnamefont{T.}~\bibnamefont{Takahashi}},
  \bibinfo{author}{\bibfnamefont{H.}~\bibnamefont{Yui}}, \bibnamefont{and}
  \bibinfo{author}{\bibfnamefont{T.}~\bibnamefont{Sawada}},
  \bibinfo{journal}{J. Phys. Chem. B} \textbf{\bibinfo{volume}{106}},
  \bibinfo{pages}{2314} (\bibinfo{year}{2002}).

\bibitem[{\citenamefont{Kovalchuk and Vollhardt}(2004)}]{kovalchulk04}
\bibinfo{author}{\bibfnamefont{N.~M.} \bibnamefont{Kovalchuk}}
  \bibnamefont{and}
  \bibinfo{author}{\bibfnamefont{D.}~\bibnamefont{Vollhardt}},
  \bibinfo{journal}{Phys. Rev. E} \textbf{\bibinfo{volume}{69}},
  \bibinfo{pages}{016307} (\bibinfo{year}{2004}).

\bibitem[{\citenamefont{Magome and Yoshikawa}(1996)}]{magome96}
\bibinfo{author}{\bibfnamefont{N.}~\bibnamefont{Magome}} \bibnamefont{and}
  \bibinfo{author}{\bibfnamefont{K.}~\bibnamefont{Yoshikawa}},
  \bibinfo{journal}{J. Phys. Chem.} \textbf{\bibinfo{volume}{100}},
  \bibinfo{pages}{19102} (\bibinfo{year}{1996}).

\bibitem[{\citenamefont{Shioi et~al.}(2003)\citenamefont{Shioi, Katano, and
  Onodera}}]{shioi03}
\bibinfo{author}{\bibfnamefont{A.}~\bibnamefont{Shioi}},
  \bibinfo{author}{\bibfnamefont{K.}~\bibnamefont{Katano}}, \bibnamefont{and}
  \bibinfo{author}{\bibfnamefont{Y.}~\bibnamefont{Onodera}},
  \bibinfo{journal}{J. Colloid Interface Sci.} \textbf{\bibinfo{volume}{266}},
  \bibinfo{pages}{415} (\bibinfo{year}{2003}).

\bibitem[{\citenamefont{Yoshikawa et~al.}(1986)\citenamefont{Yoshikawa, Omochi,
  Matsubara, and Kourai}}]{yoshikawa_omochi86}
\bibinfo{author}{\bibfnamefont{K.}~\bibnamefont{Yoshikawa}},
  \bibinfo{author}{\bibfnamefont{T.}~\bibnamefont{Omochi}},
  \bibinfo{author}{\bibfnamefont{Y.}~\bibnamefont{Matsubara}},
  \bibnamefont{and} \bibinfo{author}{\bibfnamefont{H.}~\bibnamefont{Kourai}},
  \bibinfo{journal}{Biophys. Chem.} \textbf{\bibinfo{volume}{24}},
  \bibinfo{pages}{111} (\bibinfo{year}{1986}).

\bibitem[{\citenamefont{Greenspan}(1978)}]{greenspan78}
\bibinfo{author}{\bibfnamefont{H.~P.} \bibnamefont{Greenspan}},
  \bibinfo{journal}{J. Fluid Mech.} \textbf{\bibinfo{volume}{84}},
  \bibinfo{pages}{125} (\bibinfo{year}{1978}).

\bibitem[{\citenamefont{de~Gennes}(1985)}]{degenne85}
\bibinfo{author}{\bibfnamefont{P.~G.} \bibnamefont{de~Gennes}},
  \bibinfo{journal}{Rev. Mod. Phys.} \textbf{\bibinfo{volume}{57}},
  \bibinfo{pages}{827} (\bibinfo{year}{1985}).

\bibitem[{\citenamefont{Brochard}(1989)}]{brochard89}
\bibinfo{author}{\bibfnamefont{F.}~\bibnamefont{Brochard}},
  \bibinfo{journal}{Langmuir} \textbf{\bibinfo{volume}{5}},
  \bibinfo{pages}{432} (\bibinfo{year}{1989}).

\bibitem[{\citenamefont{de~Gennes et~al.}(2004)\citenamefont{de~Gennes,
  Brochard, and Qu\'er\'e}}]{degennecap04}
\bibinfo{author}{\bibfnamefont{P.~G.} \bibnamefont{de~Gennes}},
  \bibinfo{author}{\bibfnamefont{F.}~\bibnamefont{Brochard}}, \bibnamefont{and}
  \bibinfo{author}{\bibfnamefont{D.}~\bibnamefont{Qu\'er\'e}},
  \emph{\bibinfo{title}{Capillarity and Wetting Phenomena Drops, Bubbles,
  Pearls, Waves.}} (\bibinfo{publisher}{Springer-Verlag},
  \bibinfo{year}{2004}).

\bibitem[{\citenamefont{de~Gennes}(1998)}]{degenne98}
\bibinfo{author}{\bibfnamefont{P.~G.} \bibnamefont{de~Gennes}},
  \bibinfo{journal}{Physica A} \textbf{\bibinfo{volume}{249}},
  \bibinfo{pages}{196} (\bibinfo{year}{1998}).

\bibitem[{\citenamefont{Bain and Whitesides}(1989)}]{bain89}
\bibinfo{author}{\bibfnamefont{C.~D.} \bibnamefont{Bain}} \bibnamefont{and}
  \bibinfo{author}{\bibfnamefont{G.~M.} \bibnamefont{Whitesides}},
  \bibinfo{journal}{Langmuir} \textbf{\bibinfo{volume}{5}},
  \bibinfo{pages}{1370} (\bibinfo{year}{1989}).

\bibitem[{\citenamefont{Chaudhury and Whitesides}(1992)}]{chaudhury92}
\bibinfo{author}{\bibfnamefont{M.~K.} \bibnamefont{Chaudhury}}
  \bibnamefont{and} \bibinfo{author}{\bibfnamefont{G.~M.}
  \bibnamefont{Whitesides}}, \bibinfo{journal}{Science}
  \textbf{\bibinfo{volume}{256}}, \bibinfo{pages}{1539} (\bibinfo{year}{1992}).

\bibitem[{\citenamefont{Bain et~al.}(1994)\citenamefont{Bain, Burnetthall, and
  Montgomerie}}]{bain94}
\bibinfo{author}{\bibfnamefont{C.~D.} \bibnamefont{Bain}},
  \bibinfo{author}{\bibfnamefont{G.~D.} \bibnamefont{Burnetthall}},
  \bibnamefont{and} \bibinfo{author}{\bibfnamefont{R.~R.}
  \bibnamefont{Montgomerie}}, \bibinfo{journal}{Nature}
  \textbf{\bibinfo{volume}{372}}, \bibinfo{pages}{414} (\bibinfo{year}{1994}).

\bibitem[{\citenamefont{Santos and Ondar\c{c}uhu}(1995)}]{santos95}
\bibinfo{author}{\bibfnamefont{F.~Domingues~Dos} \bibnamefont{Santos}}
  \bibnamefont{and}
  \bibinfo{author}{\bibfnamefont{T.}~\bibnamefont{Ondar\c{c}uhu}},
  \bibinfo{journal}{Phys. Rev. Lett.} \textbf{\bibinfo{volume}{75}},
  \bibinfo{pages}{2972} (\bibinfo{year}{1995}).

\bibitem[{\citenamefont{Lee and Laibinis}(2000)}]{slee00}
\bibinfo{author}{\bibfnamefont{S.}~\bibnamefont{Lee}} \bibnamefont{and}
  \bibinfo{author}{\bibfnamefont{P.~E.} \bibnamefont{Laibinis}},
  \bibinfo{journal}{J. Am. Chem. Soc.} \textbf{\bibinfo{volume}{122}},
  \bibinfo{pages}{5395} (\bibinfo{year}{2000}).

\bibitem[{\citenamefont{Bico and Qu\'er\'e}(2000)}]{bico00}
\bibinfo{author}{\bibfnamefont{J.}~\bibnamefont{Bico}} \bibnamefont{and}
  \bibinfo{author}{\bibfnamefont{D.}~\bibnamefont{Qu\'er\'e}},
  \bibinfo{journal}{Europhys. Lett.} \textbf{\bibinfo{volume}{51}},
  \bibinfo{pages}{546} (\bibinfo{year}{2000}).

\bibitem[{\citenamefont{Lee et~al.}(2002)\citenamefont{Lee, Kwok, and
  Laibinis}}]{slee02}
\bibinfo{author}{\bibfnamefont{S.~W.} \bibnamefont{Lee}},
  \bibinfo{author}{\bibfnamefont{D.~Y.} \bibnamefont{Kwok}}, \bibnamefont{and}
  \bibinfo{author}{\bibfnamefont{P.~E.} \bibnamefont{Laibinis}},
  \bibinfo{journal}{Phys. Rev. E} \textbf{\bibinfo{volume}{65}},
  \bibinfo{pages}{051602} (\bibinfo{year}{2002}).

\bibitem[{mov()}]{movies}
\bibinfo{note}{See EPAPS Document for movies of irregular motion (movie1.mpg),
  shuttling motion (movie2.mpg), rotating motion (movie3.mpg) and climbing
  motion on stair-like substrate (movie4.mpg). Every movie shown is in half
  speed of its real motion. Detailed setup is explained in suminomovietext.doc,
  which is also in EPAPS Document. A direct link to this document may be found
  in the online articlefs HTML reference section. The document may also be
  reached via the EPAPS homepage (http://www.aip.org/pubservs/epaps.html) or
  from ftp.aip.org in the directory /epaps/. See the EPAPS homepagefor more
  information.}

\bibitem[{sim()}]{simulation}
\bibinfo{note}{Simulation was performed through fourth-order Runge-Kutta method
  with the time step of 0.0001. Each parameter was taken to be as following:
  $\eta/m=24$, $\epsilon/m=0.9$, $\sigma/m = 10$, $d = 0.001$, $a=0.013$ and
  (coefficient of rebound)=0.5.}

\bibitem[{one()}]{one_d}
\bibinfo{note}{Additional explanation on the simulation, in Figure 3: The
  mechanism to generate quasi one-dimensional motion on a narrow stripe. From
  Eq. (1), the characteristic velocity $v_{c}$ and the relaxation time
  $\tau_{r}$ are given as follows: $v_{c}=\epsilon/\eta$, $\tau_{r}=m/\eta$.
  The characteristic time span $\tau_{c}$ between each rebound is
  $\tau_{c}=d/v_{c}=\eta d/\epsilon$ where $d$ is the free width of glass
  substrate. We have concluded that, when $\tau_{c} \ll \tau_{r}$, i.e. when
  $d$ is small enough, the velocity that is perpendicular to long axis of
  substrate is suppressed as long as there is loss of velocity at each
  rebound.}

\bibitem[{for()}]{force}
\bibinfo{note}{The magnitude of the driving force and the difference in surface
  tension. In Fig.\ref{fig4}A, an oil droplet moves up vertically on a glass
  substrate. In this case, the driving force generated by the difference in
  surface tension, $\Delta\gamma$, between the rear and front should be greater
  than the gravitational force, $f_{\mathrm{grav}}$, acting on the oil droplet.
  For the experiment in Fig.\ref{fig4}A, $w\Delta\gamma = f_{\mathrm{grav}}
  \geq 0.04$ [mN], where $w$ is the width of the oil droplet. Since $w \approx
  3.4$ [mm], $\Delta\gamma$ should be larger than 11 [mN/m], which is the same
  order as acquired value in Fig.\ref{fig2}A.}

\bibitem[{\citenamefont{Schliwa and Woehlke}(2003)}]{schliwa03}
\bibinfo{author}{\bibfnamefont{M.}~\bibnamefont{Schliwa}} \bibnamefont{and}
  \bibinfo{author}{\bibfnamefont{G.}~\bibnamefont{Woehlke}},
  \bibinfo{journal}{Nature} \textbf{\bibinfo{volume}{422}},
  \bibinfo{pages}{759} (\bibinfo{year}{2003}).

\bibitem[{\citenamefont{Endow and Higuchi}(2000)}]{endow_higuchi00}
\bibinfo{author}{\bibfnamefont{S.~A.} \bibnamefont{Endow}} \bibnamefont{and}
  \bibinfo{author}{\bibfnamefont{H.}~\bibnamefont{Higuchi}},
  \bibinfo{journal}{Nature} \textbf{\bibinfo{volume}{406}},
  \bibinfo{pages}{913} (\bibinfo{year}{2000}).

\bibitem[{\citenamefont{Okuda and Hirokawa}(2000)}]{okuda_hirokawa00}
\bibinfo{author}{\bibfnamefont{Y.}~\bibnamefont{Okuda}} \bibnamefont{and}
  \bibinfo{author}{\bibfnamefont{N.}~\bibnamefont{Hirokawa}},
  \bibinfo{journal}{Proc. Natl. Acad. Sci. USA} \textbf{\bibinfo{volume}{97}},
  \bibinfo{pages}{640} (\bibinfo{year}{2000}).

\bibitem[{\citenamefont{Inoue et~al.}(2001)\citenamefont{Inoue, Iwane, Miyai,
  Muto, and Yanagida}}]{inoue_yanagida01}
\bibinfo{author}{\bibfnamefont{Y.}~\bibnamefont{Inoue}},
  \bibinfo{author}{\bibfnamefont{A.~H.} \bibnamefont{Iwane}},
  \bibinfo{author}{\bibfnamefont{T.}~\bibnamefont{Miyai}},
  \bibinfo{author}{\bibfnamefont{E.}~\bibnamefont{Muto}}, \bibnamefont{and}
  \bibinfo{author}{\bibfnamefont{T.}~\bibnamefont{Yanagida}},
  \bibinfo{journal}{Biophys. J.} \textbf{\bibinfo{volume}{81}},
  \bibinfo{pages}{2838} (\bibinfo{year}{2001}).

\end{thebibliography}
\end{document}